\documentclass[twocolumn,aps,prb,groupedaddress,showpacs]{revtex4}
\usepackage{mathrsfs}
\usepackage{bm}
\usepackage{multirow}
\usepackage{amsmath,amsfonts,amssymb}
\usepackage{array,booktabs}
\usepackage{graphicx,times,graphics,color,epsfig}

\begin{document}
\draft
\title{Numerical study of parametric pumping current in mesoscopic systems in the presence of magnetic field}

\author{Fuming Xu$^1$, Yanxia Xing$^{1,2}$, and Jian Wang$^1$}
\email{jianwang@hku.hk}
\affiliation {$^1$Department of Physics and
the Center of Theoretical and Computational Physics, The University
of Hong Kong,
Hong Kong, China. \\
$^2$Department of Physics, Beijing Institute of Technology, Beijing
100081, China. \\}

\begin{abstract}
We numerically study the parametric pumped current when magnetic
field is applied both in the adiabatic and non-adiabatic regimes. In
particular, we investigate the nature of pumped current for systems
with resonance as well as anti-resonance. It is found that in the
adiabatic regime, the pumped current changes sign across the sharp
resonance with long lifetime while the non-adiabatic pumped current
at finite frequency does not. When the lifetime of resonant level is
short, the behaviors of adiabatic and non-adiabatic pumped current
are similar with sign changes. Our results show that at the energy
where complete transmission occurs the adiabatic pumped current is
zero while non-adiabatic pumped current is non-zero. Different from
the resonant case, both adiabatic and non-adiabatic pumped current
are zero at anti-resonance with complete reflection. We also
investigate the pumped current when the other system parameters such
as magnetic field, pumped frequency, and pumping potentials.
Interesting behaviors are revealed. Finally, we study the symmetry
relation of pumped current for several systems with different
spatial symmetry upon reversal of magnetic field. Different from the
previous theoretical prediction, we find that a system with general
inversion symmetry can pump out a finite current in the adiabatic
regime. At small magnetic field, the pumped current has an
approximate relation $I(B) \approx I(-B)$ both in adiabatic and
non-adiabatic regimes.
\end{abstract}


\maketitle

\section{introduction}

The idea of parametric electron pump was first addressed by
Thouless\cite{thouless}, which is a mechanism that at zero bias a dc
current is pumped out by periodically varying two or more system
parameters. Over the years, there has been intensive research
interest concentrated on parametric electron
pump.\cite{thouless1,niu,nazarov,shutenko,chu} Electron pump has
been realized on quantum dot setup\cite{aleiner1} consisting of
AlGaAs/GaAs heterojunction\cite{switkes}. Low dimensional
nanostructures, such as carbon nanotubes (CNT)\cite{ydwei2,levitov}
and graphene\cite{graphene1,graphene2} were also proposed as
potential candidates. Investigation on electron pump also triggers
the proposal of spin pump\cite{watson,benjamin,chao}, in which a
spin current is induced by various means.

At low pumping frequency limit, the variation of the system is
relatively slow than the process of energy
relaxation\cite{switkes1}. Hence the system is nearly in equilibrium
and we could deal with the adiabatic pump by equilibrium methods. On
the other hand, non-adiabatic pump refers to the case that pumping
process is operated at a finite frequency. In the non-adiabatic
regime, non-equilibrium transport theory should be employed.
Theoretical methods adopted in the research field includes
conventional scattering matrix
theory\cite{brouwer1,brouwer2,andreev}, Floquet scattering
matrix\cite{kim1,buttiker1} and non-equilibrium Green's function
(NEGF) method\cite{baigeng1,baigeng2}, as well as other
methodologies to both adiabatic \cite{aharony} and
non-adiabatic\cite{sen} electron pumps.

The electron pump is a phase coherent phenomenon, since the cyclic
variation of system parameters affects the phase of wave function
with respect to its initial value\cite{zhou}. As a result, it is
very sensitive to the external magnetic field. In the experimental
work of Switkes \textit{et.al}\cite{switkes}, at adiabatic limit,
the pumped current of a open quantum dot system with certain spatial
symmetry is showed invariant upon reversal magnetic field. The
conclusion was confirmed by theoretical works using Floquet
scattering matrix method\cite{aleiner2,kim,buttiker2}. Later it was
numerically suggested\cite{csli} that the pumped spin current also
has certain spatial symmetries.

\begin{figure}[tp]
\centering
\includegraphics[width=\columnwidth]{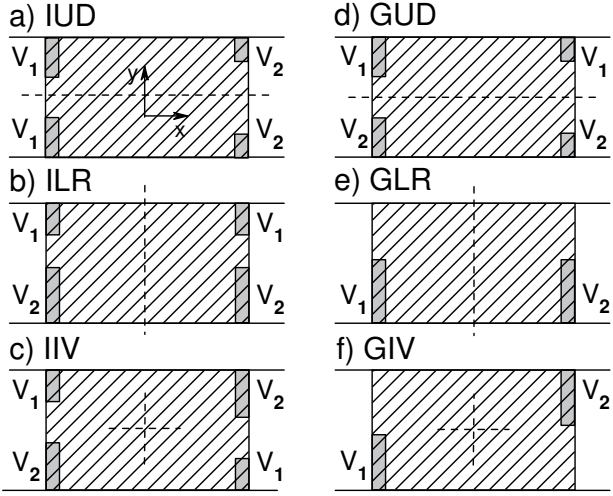}
\caption{Sketch of the spatial reflection symmetries: (a)
instantaneous up-down (IUD); (b) instantaneous left-right (ILR);
(c)instantaneous inversion (IIV); (d) general up-down (GUD); (e)
general left-right (GLR); (f) general inversion (GIV). Shadow
rectangular indicates for the pumping region and dark gray blocks
stand for potential barriers defining the spatial symmetry of the
system. The pumping potentials are right on top of these confining
potentials.}\label{fig1}
\end{figure}

In this paper, we aim to numerically investigate the pumped current
in the presence of magnetic field. Both adiabatic and non-adiabatic
pumped current are calculated. We focus on the nature of pumped
current for the mesoscopic systems with resonance with complete
transmission and anti-resonance with complete reflection. We find
that the behaviors of adiabatic and non-adiabatic pumped current are
very different. In the non-adiabatic regime, the pumped current is
nonzero at resonance while it is zero at anti-resonance. However,
the adiabatic pumped current is always zero regardless of types of
resonance. Since there is no external driving force, the direction
of current depend only on the system parameters. Our numerical
results show that the adiabatic pumped current reverses it sign at
the resonance or anti-resonance. For non-adiabatic pumped current,
the sign reversal depends on the lifetime of the resonant states.
The non-adiabatic pumped current change the sign near the resonant
point only when the lifetime is short. We also study the pumped
current as a function of magnetic field. We find that as the system
enters the quantum Hall regime with increasing magnetic field
strength, the pumping current vanishes. Since in quantum Hall regime
electron wave function appears as edge state, it will circumvent the
confining potentials shown in Fig.\ref{fig1}. Pumping potentials
overlapping in space with confining potentials present no modulation
on the electron wave function during the variation period. Hence
there is zero pumped current in the Quantum Hall regime. We also
examine the pumped current and its relation with other system
parameters such as pumping frequency and pumping potential
amplitude. Finally we also investigate the symmetry properties of
the pumped electron current of systems with certain spatial symmetry
in the presence of magnetic field by the Green's function method.
Six spatial symmetries studied in Ref.\cite{kim} are considered,
both at the adiabatic and non-adiabatic cases, which are
instantaneous up-down (IUD), left-right (ILR), and inversion
symmetries (IIV) and the corresponding non-instantaneous/general
up-down (GUD), left-right (GLR), and inversion symmetries (GIV),
respectively (see Fig.\ref{fig1}). The electron pump is driven by
periodical modulation of potentials which share the same spatial
coordinates with the confining potentials which preserve reflection
symmetry of the system. Most of our numerical results agree with the
conclusions from Floquet scattering theory\cite{aleiner2,kim},
except for the general inversion symmetry (GIV) ( setup $f$ in
Fig.\ref{fig1} ). In contrast with the theoretical prediction that
the adiabatic pumped current $I^{ad} \approx 0$ for this spatial
symmetry, our numerical calculation shows that the pumped current is
finite and further investigation reveals that there is an
approximate symmetry relation of the current as setup $e$ at small
magnetic field, which is the experimental setup\cite{switkes}. The
conclusion suggests that the general left-right (GLR) spatial
symmetry has a rather strong impact on the pumped current, which
leads to the quite accurate relation $I(B)=I(-B)$ in the
experimental finding. The result also holds for the non-adiabatic
case.

Our paper is organized as follows. In the next section we will
describe the numerical method first followed by the numerical
results and discussions in section III. Finally, conclusion is given in section IV.

\section{theoretical formalism and methodology}

We consider a quantum dot system consisting of a coherent scattering
region and two ideal leads which connect the dot to electron
reservoirs. The whole system is placed in x-y plane and a magnetic
field is applied. The single electron Hamiltonian of the scattering
region is simply
\begin{eqnarray}
H=\frac{(\textbf{p}+ e \textbf{A}/c )^2}{2m^*}+V(x,y,t) \nonumber
\end{eqnarray}
where \textbf{A} is the vector potential of the magnetic field. Here
the magnetic field is chosen to be along z-direction with
\textbf{B}=(0, 0, B). The vector potential has only x-component in
the Landau gauge , \textbf{A}=(-By, 0, 0).
\begin{eqnarray}
H=H_0+V_p \nonumber
\end{eqnarray}
where
\begin{eqnarray}
H_0=(-i\hbar \frac{\partial}{\partial x}-\frac{e}{c} By)^2+(-i\hbar
\frac{\partial}{\partial y})^2+V_0(x,y) \nonumber
\end{eqnarray}
and $V_p$ is a time-dependent pumping potential given by
$V_p(x,y,t)=\sum_j V_{j}(x,y) \cos(\omega t + \phi_{j})$.

For the adiabatic electron pump, the average current flowing through lead
$\alpha$ due to the slow variation of system parameter $V_j$ in one
period is given by\cite{brouwer1}
\begin{eqnarray}
I_\alpha & = & \frac{1}{\tau} \int_0^\tau dt
\frac{dQ_\alpha(t)}{dt} \notag\\
         & = & \frac{q\omega}{2 \pi} \int_0^\tau dt \sum_j \frac{dN_\alpha}{dV_j}
         \frac{dV_j}{dt} \label{eq1}
\end{eqnarray}
where $\tau = 2\pi / \omega$ is the variation period of parameter
$V_j$ and $\omega$ the corresponding frequency. For simplicity we
take $\omega = 1$ in the adiabatic case. $\alpha$= L or R labels the
lead. The so-called emissivity $dN_\alpha / dV_j$ is conventionally
defined in terms of the scattering matrix $S_{\alpha \beta}$
as\cite{brouwer1,ydwei1}
\begin{eqnarray}
\frac{dN_\alpha}{dV_j}= \int \frac{dE}{2 \pi} ( -\partial_E f)
\sum_{\beta} \text{Im} \frac{\partial S_{\alpha \beta}}{\partial
V_j} S_{\alpha \beta}^{\ast} \label{dnde}
\end{eqnarray}
In the language of Green's function, the above equation is
equivalent to the following form\cite{baigeng2}
\begin{eqnarray}
I_\alpha = q \int_0^{\tau} dt \int \frac{dE}{2 \pi} ( -\partial_E f)
\text{Tr} \left[ \Gamma_{\alpha} G^r \frac{dV_p}{dt} G^a
\right]\label{eq2}
\end{eqnarray}
where the instantaneous retarded Green's function $G^r$ in
real space is defined as
\begin{eqnarray}
G^r(E,t)=(E-H(t)-\Sigma^r)^{-1}
\end{eqnarray}
where $\Sigma^r$ is the self-energy due to the leads.

Whereas for non-adiabatic pump at finite frequency, the pumped
current up to the second order in pumping potential is derived as \cite{baigeng1}
\begin{eqnarray}
&&I_\alpha  =  -iq  \sum_{jk=1,2} \int
\frac{dE}{8 \pi} \text{Tr} [ \Gamma_{\alpha} G^{r}_0 V_j((f-f_{-})(G^{r-}_0 - \notag \\
& & G^{a-}_0)e^{i \phi_{kj}}  + (f-f_{+})(G^{r+}_0-G^{a+}_0)e^{-i
\phi_{kj}}) V_k G^{a}_0  ] \label{eq3}
\end{eqnarray}
where $\Gamma_{\alpha}$ is the linewidth function of lead $\alpha$
defined as $\Gamma_{\alpha}=i[\Sigma_{\alpha}^{r}
-\Sigma_{\alpha}^{a}]$; $f=f(E)$ and $f_{\pm}=f(E \pm \omega)$ are
the Fermi distribution functions; $\phi_{kj} = \varphi_j -
\varphi_k$ is the phase difference between the two pumping
potentials. Here $G^{0r} = G^{0r}(E)$ and $G^{0r \pm} = G^{0r \pm}(E
\pm \omega)$ are the retarded Green's functions where there is no
pumping potentials. In the following section, we will use Eqs.
(\ref{eq2}) and (\ref{eq3}) to carry on numerical investigations and
all the numerical work are done at the zero temperature. In the
calculation we consider a square quantum dot with size $0.7 \mu m
\times 0.7 \mu m$, of the same order as in the experimental
setup.\cite{switkes} Two open leads with the same width connect the
dot to the electron reservoirs. The quantum dot is then discretized
into a $40 \times 40$ mesh. Hopping energy $t=\hbar^2/2 m^* a^2$
sets the energy scale with $a$ the lattice spacing and $m^*$ the
effective mass of electrons in the quantum dot. Dimensions of other
relevant quantities are then fixed with respect to $t$.

\section{numerical results and discussion}

In this section, numerical results will be presented. To test our
numerical method, we first study the reflection symmetry of pumped current on inverse
of magnetic field. Other properties of the current will be discussed
in the second sub-section.

To check the symmetry of the pumped current we assume $V_0=\sum_{j=1,2} V_j(x,y)$
in our numerical calculation. In Fig.\ref{fig1}, we have schematically plotted 6
setups with different spatial symmetries of interest. In setups $a, b$, and $c$, the
spatial symmetries are kept at any moment during the pumping
period. Hence we label them instantaneous up-down (IUD),
instantaneous left-right (ILR), and instantaneous inversion (IIV)
symmetries, respectively. On the other hand, symmetries
are broken during the whole pumping cycle except when
$\phi_{jk}=n\pi$ in setup $d, e$, and $f$. They are correspondingly
labeled as general up-down (GUD), general left-right (GLR), and
general inversion (GIV) symmetries. All potential profiles locate at
the boundary of the pumping region, i.e., the first and/or last layer
in the discrete lattice (see the dark gray region).

\subsection{Symmetry of pumped current}

Before the presenting numerical results, we would like to point out
that for setup $c$ with instantaneous inversion symmetry (IIV), the
theoretical predictions \cite{aleiner2,kim} and our numerical
calculations give the same result: the pumped current is exact zero
at both adiabatic and non-adiabatic cases that is independent of $B$
and $\phi_{12}$. The phenomena can be straightforwardly understood
by Floquet scattering matrix theory\cite{kim}. In IIV setup, it is
obvious that the transmission coefficient of an electron traveling
from the left lead to the right $T_{R \leftarrow L}$ is always equal
to that of an electron moving in the opposite direction, $T_{L
\leftarrow R}$, i.e.,
\begin{eqnarray}
T_{R \leftarrow L} = T_{L \leftarrow R} \nonumber
\end{eqnarray}
From the Landauer-B\"{u}ttiker formula, the electric current along
the left to right region is given by
\begin{eqnarray}
I_{R \leftarrow L}=\frac{2e}{h}\int dE T_{R \leftarrow L}(E) f(E)
\nonumber
\end{eqnarray}
while $I_{L \leftarrow R}$ is defined in a similar way. Then the
pumped current through the left lead is defined
as\cite{buttiker2,kim}
\begin{eqnarray}
I_L = I_{L\rightarrow R} - I_{R\rightarrow L}=0
\end{eqnarray}
The conclusion holds for any particular moment, which means that
there will be no pumped current at all. Hence in the following we
will not discuss the case of setup $c$.

\begin{figure}[tbp]
\centering
\includegraphics[width=\columnwidth]{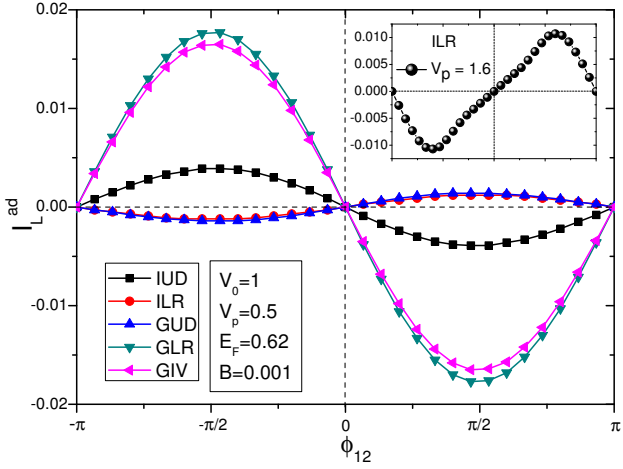}
\caption{The adiabatically pumped current as a function of phase
difference $\phi_{12}$ for different spatial symmetries of the
pumping system at pumping amplitude $V_p=0.5$. Inset: pumped current
of system with symmetry ILR at $V_p=1.6$. Other system parameters:
$E_F=0.62$, $B=0.001$, $V_0=1$. }\label{fig2}
\end{figure}

First we examine the relation between the adiabatically pumped
current $I^{ad}$ and phase difference $\phi_{12}$ of the pumping
potentials calculated from Eq.(\ref{eq2}). A sinusoidal behavior is
observed at a relatively small pumping amplitude $V_p$=0.5 for all
setups in Fig.\ref{fig1}. The sinusoidal form of $I^{ad}(\phi_{12})$ represents a
generic property of adiabatic electron pump at small $V_p$. Driven
by the cyclic variation of two time-dependent system parameters, the
pumped current is directly related to the area enclosed by the
parameters in parametric space. At small pumping amplitudes, the
leading order of $I^{ad}$ is proportional to the phase difference
between the pumping potentials, $I^{ad} \propto V_p \sin \phi_{12}$
\cite{brouwer1}. However, the relation doesn't hold for large
pumping amplitude. To demonstrate this we have calculated current
pumping through the setup with symmetry ILR at a large potential
$V_p$=1.6. As shown in Fig.\ref{fig2} the
sinusoidal relation is clearly destroyed. Except for this difference arising
from the pumping amplitude $V_p$, there is a general antisymmetry
relation between the pumped current and the phase difference
$\phi_{12}$ for all setups:
$I(\phi_{12})=-I(-\phi_{12})$. Naturally, $I(\phi_{12}=n \pi)=0$.
This is understandable since two simultaneously varying parameters enclose
a line rather than an area in the parametric space, the pumped current vanishes.
This result, however, does not hold for non-adiabatic case where the frequency
gives additional dimension of parametric space.
Another result of interest is that, in contrast to the theoretical
prediction $I^{ad} \approx 0$ for the setup of GIV symmetry \cite{kim}, the
pumped current from the setup of GIV is finite and has the same order of magnitude as that of GLR
symmetry.

\begin{figure}[tbhp]
\centering
\includegraphics[width=\columnwidth]{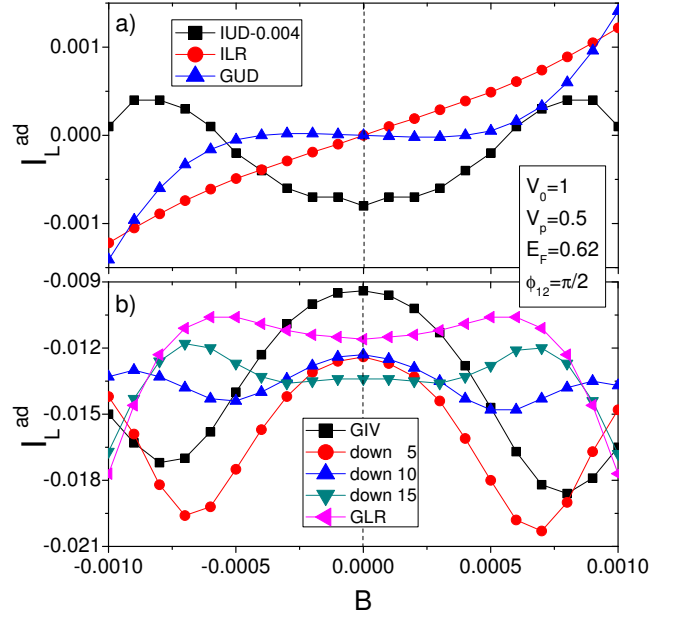}
\caption{Panel (a): The adiabatically pumped current versus magnetic
field strength $B$ for system symmetries $a$(IUD), $b$(ILR), and
$d$(GUD). Curve for IUD is offset by -0.004 for a compact
illustration. Panel (b): $I^{ad}$ vs $B$ for spatial symmetry GLR,
GIV, and three intermediate setups. Calculation parameters:
$E_F=0.62$, $\phi_{12}=\pi/2$, $V_0=1$, $V_p=0.5$. }\label{fig3}
\end{figure}

Fig.\ref{fig3} plots the pumped current versus magnetic
field strength $B$ at phase difference $\phi_{12}= \pi/2$, where the
magnitude of $I^{ad}(\phi_{12})$ is maximized. In the upper panel
(a) of Fig.\ref{fig3}, we see that current is either even or odd
function of magnetic field strength $B$ for symmetries IUD, ILR, and
GUD, $I(B) = \pm I(-B)$ which agree with the theoretical
predictions.\cite{aleiner2,kim} In panel (b), it is clear that the pumped
current is invariant upon the reversal of magnetic field for
GLR.\cite{aleiner2,kim} The system
with GIV symmetry shows an approximation relation $I(B) \approx
I(-B)$ only at small $B$, which is similar to that of GLR
symmetry. This does not agree with the theoretical prediction.\cite{kim}
To further investigate the relation $I(B) \approx
I(-B)$ we studied three intermediate
setups between GIV and GLR. In panels (e) and (f) of Fig.\ref{fig1}, the
length of the pumped potential profile is fixed as 20 both for $V_1$ and
$V_2$ in the system with $40 \times 40$ mesh. Then we shift down
the potential profile $V_2$ of the
GIV symmetry 5 lattice spacing each time. After 4 shifts, the system
changes from GIV to GLR symmetry and in this process three
intermediate systems are generated. Numerical results shown in
panel (b) of Fig.\ref{fig3} suggest that all these setups have the
relation $I(B) \approx I(-B)$ at small magnetic fields, although there
are no spatial reflection symmetries in these systems. The
closer system to the GLR symmetry, the larger the magnetic field
for this relation. These results
suggest that the general left-right symmetry is a rather strong spatial
symmetry and even a rough setup (cyan-down-triangle curve in panel
(b)) can lead to an accurate invariant relation of pumped current, at
least for small magnetic field.
Recall the experimental result \cite{switkes} of an adiabatic pump
where the experimental setup can not have precise GLR symmetry,
but an accurate relation $I(B) = I(-B)$ was still
achieved. In addition, the amplitude of pumped current in this setup
with GLR symmetry is relatively high, compared with other symmetries.

\begin{figure}[tbp]
\centering
\includegraphics[width=\columnwidth]{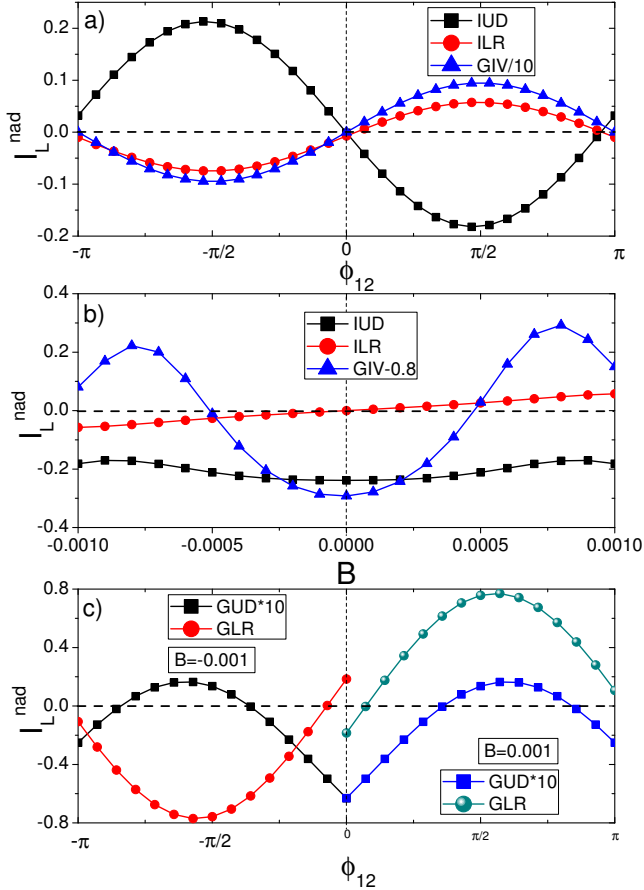}
\caption{Panel (a): Non-adiabatically pumped current as a function of
phase difference $\phi_{12}$ at a fixed magnetic field $B=0.001$ for
spatial symmetries IUD, ILR, and GIV. Blue curve with up-triangle
for GIV is multiplied by a factor of $0.1$. Panel (b): $I^{nad}$
versus magnetic field strength $B$ for the above three setups with
$\phi_{12} = \pi/2$. Blue curve with up-triangle for GIV is offset
by $-0.8$. Panel (c): The pumped current for system setups GUD and
GLR. Curves for GUD are multiplied by $0.1$. Calculation parameters:
$E_F=0.62$, $V_0=1$, $V_p=40$. }\label{fig4}
\end{figure}

Now we turn our attention to the non-adiabatic electron pump with
finite pumping frequency. The numerical results are presented in
Fig.\ref{fig4}. One of the major differences between adiabatic and
non-adiabatic pump is that a non-adiabatic pump can operate with
only one system parameter, since the finite pumping frequency
supplies one extra degree of freedom and it could act as another
pumping parameter. In the experiment\cite{switkes} it was found that
$I(\phi_{12}=0) \neq 0$. Later a theoretical work\cite{baigeng1}
attributed this phenomenon as a consequence of photon-assisted
processes and it is a nonlinear transport feature of non-adiabatic
electron pump. In our numerical results, we also found that
$I^{nad}(\phi_{12}= n\pi) \neq 0$ is a general property of the
pumped current, except for systems with spatial symmetries IIV or
GIV. Although the pumping frequency $\omega$ can play the role of a
variation parameter, the pumped current in the system with symmetry
IIV is always zero. For the setup with GIV symmetry, we see from
panel (a) of Fig.\ref{fig4} that the pumped current obeys an
antisymmetric relation with phase difference: $I^{nad}(\phi_{12})=
-I^{nad}(-\phi_{12})$. $I^{nad}(\phi_{12}= n\pi) = 0$ is a natural
result of this antisymmetry relation. Combining with the result from
the adiabatic case (Fig.\ref{fig2}), we see that this antisymmetry
relation between pumped current and phase difference $\phi_{12}$ is
a general feature of the GIV symmetry. Besides, from panel (b) of
Fig.\ref{fig4} we found that, $I^{nad}$ for GIV system at a fixed
phase $\phi_{12}= \pi /2$ shows $I^{nad}(B) \approx I^{nad}(-B)$ at
small magnetic field. This approximate symmetry relation can not be
obtained theoretically. Our results confirm the theoretical
predictions on the parity of pumped current on reversal of magnetic
field for setups IUD and ILR\cite{kim}, which are respectively
$I(B)=I(-B)$ and $I(B)=-I(-B)$ (see panel (b)). However, it doesn't
hold for GUD and GLR in panel (c). In this case, one can only get
the relations $I(B,\phi)=I(-B,-\phi)$ for GUD and
$I(B,\phi)=-I(-B,-\phi)$ for GLR\cite{kim}. When the two pumping
potentials operate in phase or out of phase ($\phi_{12} = n\pi$),
they reduce to a simple version: $I(B)=I(-B)$ for GUD and
$I(B)=-I(-B)$ for GLR, which are the same for IUD and ILR at
$\phi_{12} = n\pi$.  It is worth
mentioning that these two relations are in contrary to the adiabatic
case where $\phi_{12} \neq n\pi$.

We collect the results and summarize these conclusions drawn from
both adiabatic and nonadiabatic pumps and there are shown in
Table.\ref{table1} in detail.

\begin{table*}
\begin{tabular}{|c|c|c|c|c|}
\hline
 & \multicolumn{2}{c}{adiabatic pump} &  \multicolumn{2}{c}{nonadiabatic pump} \\ \hline
 & $\phi_{12}= n\pi$ & $\phi_{12}\neq n\pi$ & $\phi_{12}= n\pi$ & $\phi_{12}\neq n\pi$ \\ \hline
IUD & ${I=0}^{\diamond}$ & ${I(B)=I(-B)}^{\diamond}$  &
${I(B)=I(-B)}^{\diamond}$  &  ${I(B)=I(-B)}^{\diamond}$ \\ \hline
ILR & ${I=0}^{\diamond}$ & ${I(B)=-I(-B)}^{\diamond}$ &
${I(B)=-I(-B)}^{\diamond}$ &  ${I(B)=-I(-B)}^{\diamond}$ \\ \hline
IIV & ${I=0}^{\diamond}$ & ${I=0}^{\diamond}$         &
${I=0}^{\diamond}$         &  ${I=0}^{\diamond}$ \\ \hline GUD &
${I=0}^{\diamond}$ & ${I(B)=-I(-B)}^{\diamond}$ &
${I(B)=I(-B)}^{\diamond}$  &  ${I(B,\phi)=I(-B,-\phi)}^{\diamond}$
\\ \hline GLR & ${I=0}^{\diamond}$ & ${I(B)=I(-B)}^{\diamond}$  &
${I(B)=-I(-B)}^{\diamond}$ &  ${I(B,\phi)=-I(-B,-\phi)}^{\diamond}$
\\ \hline \multirow{2}{*}{GIV} &  ${I=0}^{\diamond}$   &
${I \approx 0}^{\triangleright}$  & ${I=0}^{\diamond}$ &
${I(B,\phi)=-I(B,-\phi)}^{\diamond}$ \\ & & ${I(B) \approx
I(-B)}^{\ast}$  & &  ${I(B) \approx I(-B)}^{\ast}$ \\ \hline
\end{tabular}
\caption{Symmetry of the pumped currents on inversion of the
magnetic field for both adiabatic and nonadiabatic electron pumps. \\
$\quad$ $\diamond$ stands for the theoretical prediction
from Ref.\onlinecite{kim} confirmed by our numerical calculation. \\
$\triangleright$ represents theoretical relation without numerical sustainment. \\
$\ast$ corresponds to our new finding in contrast to the theoretical
prediction. } \label{table1}
\end{table*}

\subsection{Transport properties of the pumped current}

\begin{figure}[tbhp]
\centering
\includegraphics[width=\columnwidth]{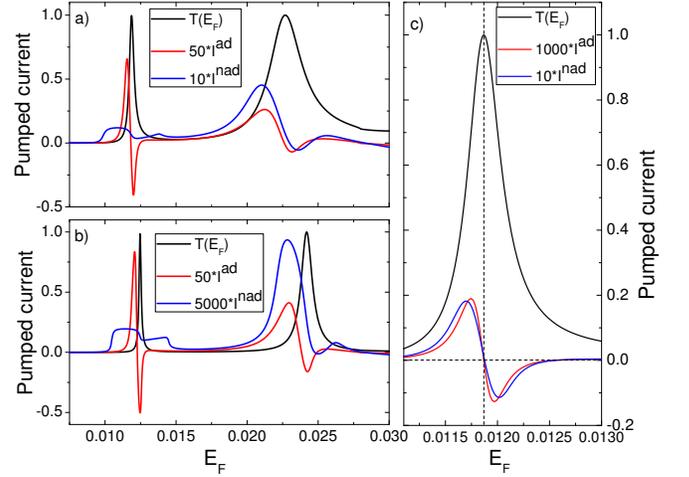}
\caption{Panel (a) and (b): The pumped current as well as
transmission coefficient as a function of Fermi energy at static
potential height $V_0=1.0$ and $V_0=5.0$, respectively. For
visualization purpose, a factor is multiplied to the pumped current
in Fig.\ref{fig5}. For $I^{ad}$ the factor is 50 in both panels. For
$I^{nad}$ this factor is 10 in panel (a) and 5000 in panel (b).
Other parameters: $B=0.001$, $\phi_{12}=\pi/2$. $V_p=0.5$,
$\omega=0.002$ in panel (a) and $V_p=4.5$, $\omega=0.002$ in panel
(b). Panel (c) highlights the pumped current at small pumping
amplitudes at the first resonant peak. $V_0=1.0$ in this panel and
$V_p=0.05$, $\omega=0.0002$. The factors for $I^{ad}$ and $I^{nad}$
are 1000 and 10, respectively. }\label{fig5}
\end{figure}

\begin{figure}[tbhp]
\centering
\includegraphics[width=\columnwidth]{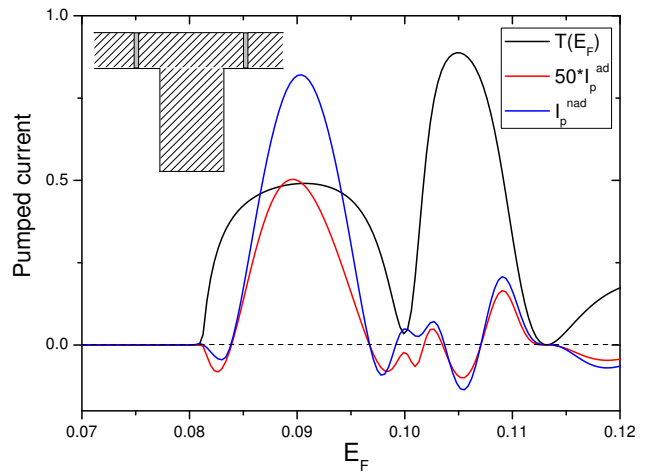}
\caption{The pumped current versus Fermi energy in a T-shaped
system. The side bar is of length 20 and width 30. Two gray blocks
indicate the positions where the pumping potentials are applied and
the static potential is set to be zero. A factor of 50 is multiplied
to $I^{ad}$. Other parameters: $B=0.001$, $\phi_{12}=\pi/2$.
$V_p=0.05$, $\omega=0.002$.}\label{fig6}
\end{figure}

\begin{figure}[tbhp]
\centering
\includegraphics[width=\columnwidth]{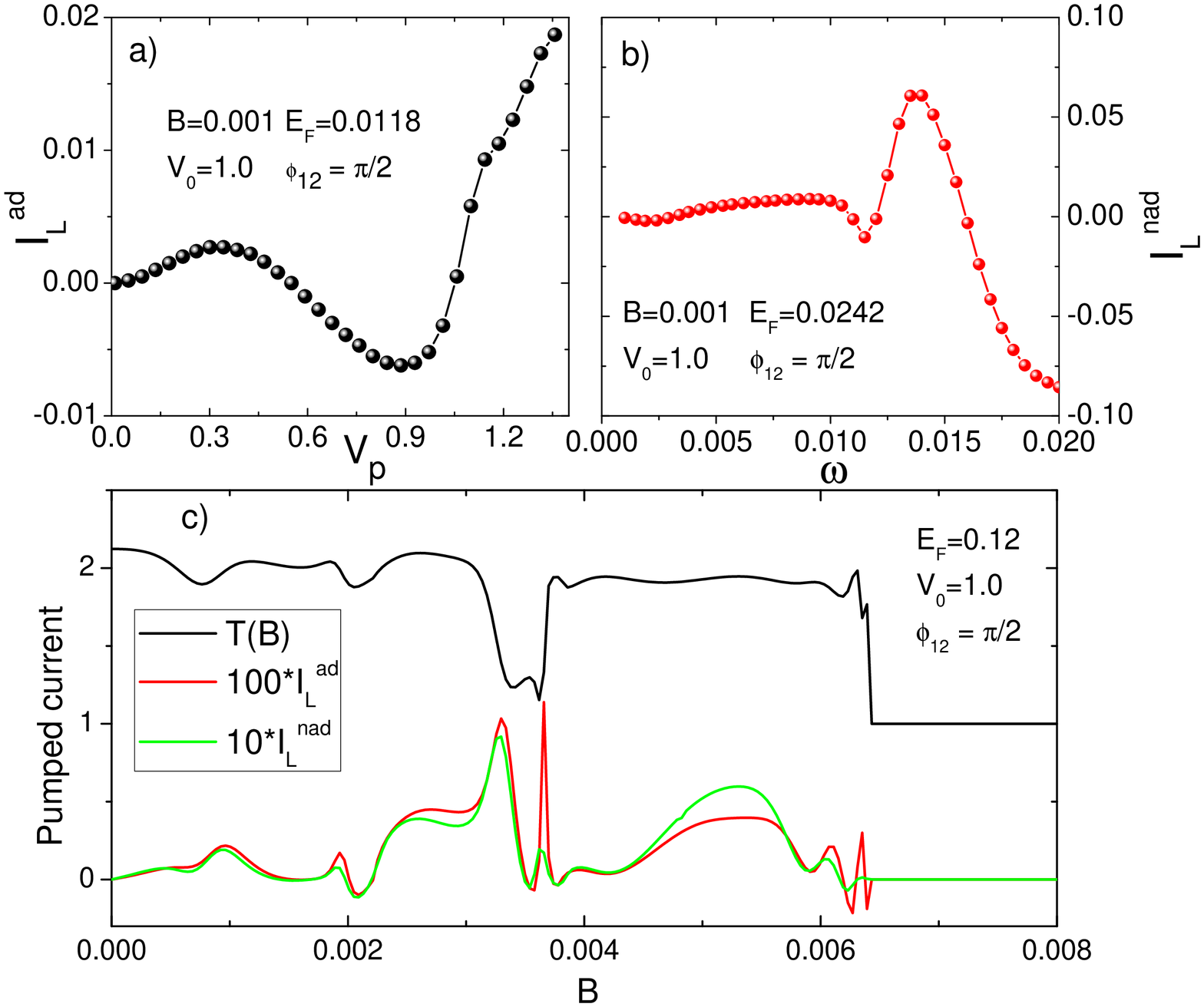}
\caption{Panel (a): Adiabatic current vs pumping potential $V_p$ at
$E_F=0.0118$. Panel (b): Non-adiabatic current versus pumping
frequency $\omega$ at $E_F=0.0242$. System parameters:
$\phi_{12}=\pi /2$, $V_0=1$, $B=0.001$. Panel (c) shows the pumped
current and transmission coefficient versus magnetic field $B$ at
Fermi energy $E_F=0.12$, with other parameters: $\phi_{12}=\pi /2$,
$V_0=1$, $V_p=0.5$, $\omega=0.002$. For illustration a factor of 100
is multiplied to $I^{ad}$ and it is 10 for $I^{nad}$. }\label{fig7}
\end{figure}

In the last section we have concentrated on the symmetry of the
pumped current with magnetic field $B$ and phase difference
$\phi_{12}$ as the variables. Now we study the effect of other
system parameters on the pumped current. Numerical calculations were
performed on a system with instant L-R symmetry (ILR), in which
widths of the four potential barriers are kept equal. The numerical
results are plotted in Fig.\ref{fig5}, Fig.\ref{fig6} and
Fig.\ref{fig7}.

In panel (a) of Fig.\ref{fig5} we plot the pumped current in the
presence of magnetic field as a function of Fermi energy $E_F$,
together with the transmission coefficient $T(E_F)$ at static
potential barrier $V_0=1.0$. The sharp tips of transmission
coefficient suggests that quantum resonance effect dominates the
transport process. When a dc bias is applied, the tunneling current
is calculated from transmission profile. However, the pumped current
is generated as zero bias by periodically varying ac gate voltages.
Although originating from different physical mechanisms, we see that
the pumped current clearly show resonance characteristics both in
adiabatic and non-adiabatic cases near the resonant energy of static
transmission coefficient. These resonance-assisted behavior of
pumped current is a generic property of electron pump.\cite{ydwei1}
Operating at the coherent regime, quantum interference naturally
results in its resonant behavior. It is worth mentioning that near
the sharp resonance at $E_F=0.0118$ the adiabatic pumped current
changes sign. This is understandable. In the presence of dc bias,
the direction of the current is determined by the bias. For
parametric electron pump at zero bias, the direction of the pumped
current depends only on the system parameters such as Fermi energy
and magnetic field. Variation of these parameters can change the
current direction. For the non-adiabatic pump, the pumped current
changes slowly near the resonance but there is no sign changes for
the pumped current. In Fig.\ref{fig5}(a), we also see a second
resonant point with much broader peak. Near this resonant level, we
see that the transmission coefficient and the pumped currents are
well correlated. The resonant feature of the pumped current is also
related to the width of the resonant peak in the transmission
coefficient. Similar behaviors are found for a higher static
potential barrier $V_0=5.0$ in panel (b). A larger barrier makes the
resonant peaks much sharper, but it doesn't qualitatively affect the
pumped current. The noticeable difference is that, the pumped
current peaks are shifted with that of the transmission coefficient.
In addition, it seems that the non-adiabatic pumped current develops
a plateau region near the first resonant peak.

When zooming in at this resonant peak ($E=0.0118$ in
Fig.\ref{fig5}(a)), we found that at small
pumping amplitude the pumped currents are zeros for both adiabatic
and non-adiabatic cases when a complete transmission occurs
(transmission coefficient $T=1$). The numerical evidence is shown in
panel (c) of Fig.\ref{fig5}. Note that there is only one transmission channel
for the incident energy so that $T=1$ corresponds to complete transmission.
We emphasize that the non-adiabatic pumped
current goes to zero near the resonance only for very small frequency. At
larger frequency such as the case in Fig.\ref{fig5}(a) or (b), it is nonzero.
For the case of $I^{ad}$, it is easy to understand why it is zero
at $T=1$. For a perfect transmission, the diagonal terms $S_{LL}$ and $S_{RR}$
of the four-block scattering matrix are zero. Hence we have
$S_{LR}=S_{RL}=\exp(i\theta)$. From Eq.(\ref{dnde}), we have
\begin{eqnarray}
dN_\alpha/dV_j = (i\partial \theta/\partial V_j)/\pi \label{dnde1}
\end{eqnarray}
For two pumping potentials,
the current can be expressed in parameter space. Using the Green's theorem,
Eq.(\ref{eq1}) becomes\cite{brouwer1},
\begin{eqnarray}
I_\alpha & = & \frac{1}{\tau} \int_0^\tau dt
\frac{dQ_\alpha(t)}{dt} \notag\\
         & = & \frac{q\omega}{2 \pi} \int dV_1 dV_2 \left( \frac{\partial}{\partial
V_1}\frac{dN_\alpha}{dV_2} - \frac{\partial}{\partial
V_2}\frac{dN_\alpha}{dV_1} \right) \label{para}
\end{eqnarray}
From Eq.(\ref{dnde1}), it is easy to see that the integrand is zero.
Hence $I_\alpha=0$ if $S_{\alpha \alpha}=0$. For non-adiabatic case,
the pumped current at complete transmission is in general nonzero.
However, if the frequency is very small, the adiabatic case is
recovered. This is numerically supported by Fig.\ref{fig5}(c), where
the two current curves are very similar.

Furthermore, the behavior of pumped current for a structure exhibiting
anti-resonance phenomena was studied and the numerical results is
shown in Fig.\ref{fig6}. To establish anti-resonance, we use a
T-junction\cite{T-junction}, which is schematically
plotted in the inset of Fig.\ref{fig6}. The side bar has longitudinal dimension
20 and transverse dimension 30. Pumping potentials are placed on the
two arms of the device and there are no static potential barriers in
the system. From the transmission curve shown in Fig.\ref{fig6}, one clearly finds that
$T(E_F)$ drops sharply to zero around $E_F=0.112$, which is the
signature of anti-resonance. At this point, both the adiabatic and
nonadiabatic pumped current are zero. Different from the resonant case, here the
range where the current is zero or nearly zero is much broader.
The phenomena are attributed to reasons similar to those
we presented above, but in this case $S_{LR}$ and $S_{RL}$ are zero
at the anti-resonance point. We also see that at transmission minimum
$E_F=0.10$ with small transmission coefficient the pumped
current is nonzero.

In panel (a) and (b) of Fig.\ref{fig7}, we examine the influence of
pumping amplitude $V_p$ on the pumped currents with the static
potential barrier fixed at $V_0=1$, which corresponds to
the case shown in panel (a) of Fig.\ref{fig5}. At the first resonant
energy $E_F=0.0118$ we plot $I^{ad}$ versus $V_p$, the pumping
potential amplitude. The non-adiabatic pumped current $I^{nad}$ as
a function of the pumping
frequency $\omega$ is evaluated at the second resonant peak
$E_F=0.0242$. In both cases, magnitudes of the pumped current
changes in a oscillatory fashion with the increasing of $V_p$ or $\omega$. The
pumped current can change its sign, which also reflects the nature of the
parametric pump and manifest distinction between the pumped current
and the conventional resonant tunneling current.

\begin{figure}[tbhp]
\centering
\includegraphics[width=\columnwidth]{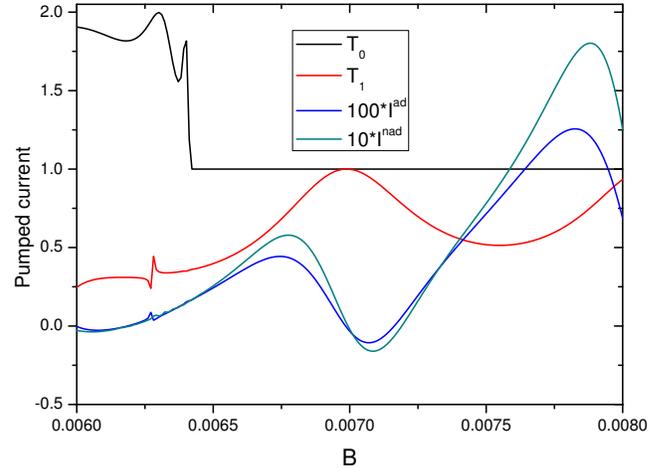}
\caption{The pumped current as well as transmission coefficient as a
function of magnetic field $B$ at Fermi energy $E_F=0.12$. Other
parameters: $\phi_{12}=\pi /2$, $V_0=1$, $V_p=0.5$, $\omega=0.002$.
$I^{ad}$ and $I^{nad}$ are scaled by factors of 100 and 10,
respectively. }\label{fig8}
\end{figure}

The resonance behavior of pumped current is also visible in panel
(c) of Fig.\ref{fig7}, in which we depict $I_p$ and transmission
coefficient versus magnetic field $B$ at Fermi energy $E_F=0.12$.
Sweeping through magnetic field, there is a sharp change of
transmission coefficient near $B \sim 0.003$ and the pumped current
changes accordingly. With increasing magnetic field, $T$ becomes
quantized (there is only one transmission channel at this magnetic
field) indicating the occurrence of edge states in the quantum Hall
regime and the pumped current vanishes. In our setup, electron pump
operates by cycling modulation of electron passing through the
pumping potentials which are on top of static barriers defining the
system. With increasing of the magnetic field, electron wavefunction
tends to localize near the edge, which decreases the modulation
efficiency of the pumping potentials. As the edge state emerges,
electron will circumvent the confining potentials with no reflection
during their deformations. In this case, the variation of pumping
potential has no effect on the moving electron. Hence there is no
pumped current when edge state is formed in the system.
Mathematically it is also easy to show from Eq.(\ref{eq1}) that for
a two-probe system as long as the instantaneous reflection
coefficient vanishes (in the case of edge state) in the whole
pumping period there is no adiabatic pumped current.

We provide a numerical evidence for the above statement, which is
shown in Fig.\ref{fig8}. In contrast to the calculation of panel (c)
of Fig.\ref{fig7}, the static potential barriers extend to a width
40, which is exactly the width of the scattering region. At the same
time, the pumping barriers remain the same as before (with width 10).
Now the static transmission coefficient, labeled $T_1$ in the figure,
does not have quantized
value but exhibits a resonant behavior. $T_0$ is copied from
Fig.\ref{fig7} for comparison. As long as the edge state of an
electron is scattered with transmitted and reflected modes, the
pumped current will be generated with varying system parameters.

\section{conclusion}

In conclusion, we have studied the pumped current as a function of
pumping potential, magnetic field and pumping frequency in the
resonant and anti-resonant tunneling regimes. Resonant features are
clearly observed for adiabatic and non-adiabatic pumped current. We
found that when the resonant peak is sharp the adiabatic pumped
current changes sign near the resonance while non-adiabatic pumped
current does not. When the resonant peak is broad the behaviors of
pumped current in adiabatic and non-adiabatic regimes are similar
and both change sign near the resonance. At anti-resonance, however,
both adiabatic and non-adiabatic pumped current are zero. As the
system enters the quantum Hall regime, pumped currents vanishes in
all the setups shown in Fig.\ref{fig1}, since the pumping potentials
can not modulate the electron wave function. Furthermore, we have
numerically investigated the symmetry of the adiabatic and
non-adiabatic pumped current of systems with different symmetries
placed in magnetic field. The calculated results are listed in
Table.\ref{table1} and most of them are in agreement with the former
theoretical results derived from Floquet scattering matrix theory.
Different from the theoretical prediction, we found that the system
with general spatial inversion symmetry (GIV) gives rise to a finite
pumped current at adiabatic regime. At small magnetic field, both
the adiabatic and non-adiabatic currents have an approximation
relation $I(B) \approx I(-B)$.

\section{acknowledgments}

This work is supported by RGC grant (HKU 705409P), University Grant
Council (Contract No. AoE/P-04/08) of the Government of HKSAR, and LuXin Energy Group.
The computational work is partially performed on HPCPOWER2 system of the computer center, HKU.

\end{document}